\documentclass[10pt,letterpaper,compsoc,conference]{iiswc24}

\usepackage{cite}
\usepackage{amsmath,amssymb,amsfonts}
\usepackage{algorithmic}
\usepackage{graphicx}
\usepackage[dvipsnames]{xcolor}
\usepackage[final]{microtype}
\usepackage[italic]{mathastext}
\usepackage{libertine}
\usepackage[T1]{fontenc}
\usepackage{textcomp}
\usepackage[varqu,varl]{zi4}
\usepackage[all]{nowidow}
\usepackage[keeplastbox]{flushend}
\usepackage{fancyhdr}

\usepackage{subcaption}
\usepackage{tabularx}
\usepackage{booktabs}
\usepackage{multirow}
\usepackage{siunitx}
\def\BibTeX{{\rm B\kern-.05em{\sc i\kern-.025em b}\kern-.08em
    T\kern-.1667em\lower.7ex\hbox{E}\kern-.125emX}}
\usepackage{pifont}

\fancypagestyle{firstpage}{
  \fancyhf{}
  
  \fancyfoot[C]{\thepage}
}

\begin{document}



\title{Performance Modeling and Workload Analysis of Distributed Large Language Model Training and Inference}





\author{\IEEEauthorblockN{Joyjit Kundu$^{\dagger \ast}$, Wenzhe Guo$^{\dagger \ast}$, Ali BanaGozar$^{\dagger \ast}$, Udari De Alwis$^{\ast}$,  Sourav Sengupta$^{\ast}$, Puneet Gupta$^{\ddagger}$ and \\ Arindam Mallik$^{\ast}$}
\IEEEauthorblockA{\textit{$^{\ast}$Interuniversity Microelectronics Centre (IMEC), Leuven, Belgium} \\
{$^{\ddagger}$Department of Electrical and Computer Engineering,
University of California, Los Angeles} \\
$\dagger$\{joyjit.kundu, wenzhe.guo\}@imec.be,~a.banagozar@tue.nl}}

\maketitle
\def\thefootnote{$\dagger$}\footnotetext{These authors contributed equally to this work}
\thispagestyle{firstpage}
\pagestyle{plain}

\begin{abstract} 
Aligning future system design with the ever-increasing compute needs of large language models (LLMs) is undoubtedly an important problem in today's world.
Here, we propose a general performance modeling methodology and workload analysis of distributed LLM training and inference through an analytical framework that accurately considers compute, memory sub-system, network, and various parallelization strategies (model parallel, data parallel, pipeline parallel, and sequence parallel).
We validate our performance predictions with published data from literature and relevant industry vendors (e.g., NVIDIA). For distributed training, we investigate the memory footprint of LLMs for different activation re-computation methods, dissect the key factors behind the massive performance gain from A100 to B200 ($\sim$ 35x speed-up closely following NVIDIA's scaling trend), and further run a design space exploration at different technology nodes (12 nm to 1 nm) to study the impact of logic, memory, and network scaling on the performance. For inference, we analyze the compute versus memory boundedness of different operations at a matrix-multiply level for different GPU systems and further explore the impact of DRAM memory technology scaling on inference latency. Utilizing our modeling framework, we reveal the evolution of performance bottlenecks for both LLM training and inference with technology scaling, thus, providing insights to design future systems for LLM training and inference. 

\end{abstract}
\section{Introduction \& Background}
\label{sec:back}
Transformer architecture has emerged as one of the most widely used neural network architectures in various artificial intelligence (AI) applications domain. The whole zoo of large language models (LLMs) (e.g., class of GPT models and Llamma variants) with their ever-increasing model size are examples of transformer architecture, playing a dominant role in today's natural language processing and image classification\cite{Yiheng2024}. Training a large language model requires a huge amount of data and compute time, resulting in significant carbon emissions along with an estimated cost of tens of thousands to millions of dollars. For instance, training a GPT-3 transformer model costs around \$10M~\cite{Ivanov2020DataMI,wiggers2020}. However, the cost in the long run can be dominated by inference when the same model is deployed to serve multiple users for a long period of time. Thus, it is important to understand the impact of the trend in LLMs and underlying system architecture on the performance per total cost of operation (TCO) in the context of both large-scale training and inference. Detailed analysis of the performance per TCO would help us identify the pain points and invest in required areas or components while designing future compute systems or models.
\subsection{Transformers}\label{subsec:Transformer}
The decoder-based transformer architecture is quite regular and each layer of it primarily consists of a multi-head attention block (MHA) and a multi-layer perceptron block (MLP)\cite{Vaswani2017}. This structural regularity and {\em almost} static nature of the data flow at a higher abstraction level allow analytical modeling of the performance of LLMs at a data center level\cite{AMPeD}. The attention mechanism is at the heart of the transformer architecture. The prediction accuracy of LLMs depends on the sequence length or the context of the model\cite{Metzler2020}. Unfortunately, the execution time and memory complexity of attention grows quadratically with sequence length\cite{flashattention}. An important challenge in the LLM community is scaling the performance of transformer models with long sequences. There are primarily three aspects that determine the performance: the number of floating point operations (FLOPs), memory accesses, and communication over the network. For instance, the FlashAttention~\cite{flashattention,flashattention2} work addresses this problem for LLM training by focusing on the memory access to and from DRAM at the cost of FLOPs. Similarly, the implementation of KV-cache is crucial for scaling the performance of inference. However, the performance bottleneck is not static and often shifts with the evolution of LLMs, compute architectures, or technology. A recent analysis suggests communication will have a significant overhead (40-75\%) in runtime as models and hardware evolve \cite{pati2023}. Thus, a generic framework that can expose different tradeoffs in the performance of LLM training and inference workloads with a connection to the technology is essential for hardware-software co-design. 
\subsection{Performance Bottlenecks}\label{subsec:perf}
The primary operations in Transformer architecture can be categorized into three groups or kernels: tensor contractions (general matrix-matrix multiply -- GEMM or matrix-vector multiply -- GEMV), normalization (e.g., softmax, layer-norm), and element-wise operations (e.g., non-linearities, biases, and dropout)~\cite{Ivanov2020DataMI}. Among these, GEMM or GEMV (depending on training or inference) is the most critical operation that dictates the overall performance of a transformer. The performance at a GEMM level can be characterized by studying the balance between pure computation and memory accesses. If the time taken by an operation is dominated by the count of arithmetic operations not the memory access time, it is categorized as {\em compute-bound}. For {\em memory-bound} operations, the execution time is primarily determined by the memory accesses while the time spent in actual computation is negligible. The arithmetic intensity is a metric that shows the compute or memory-boundedness of a kernel by capturing the number of arithmetic operations per byte transferred to and from the memory. Tensor contractions in distributed training are generally compute intensive since they involve fat GEMMs ($m\approx n \approx k >> 1$), characterized by matrices that are either square or closer to square in shape due to the substantial batch and sequence dimensions; however, in the auto-regressive generation phase of inference, they are mostly memory-bound since the inherent sequential nature of token generation leads to skinny GEMMs -characterized by matrices that are long rectangles in shape or GEMVs. The other two types of operations are in general memory-bound as well. Kernel-fusion is one of the techniques commonly used to improve the arithmetic intensity of such operations \cite{deepspeed}. The above issues concern the performance at a single accelerator level. The other important aspect of LLM training or inference at scale is the data transfer through the network. The communication overhead becomes quite important for large-scale training and inference on advanced multi-GPU systems when compute is relatively fast\cite{pati2023, deepspeed}.
\subsection{Parallelization}\label{subsec:parallelization}
Distributed training involves different parallelism strategies: Data parallelism (DP), Tensor model parallelism (TP), Pipeline parallelism, and Sequence parallelism (SP)\cite{Hoefler2019, eff_megatron, Korthikanti2022ReducingAR}. In DP, each GPU processes a portion of the data but shares the same model parameters to compute the gradients locally. The gradients across the devices are then reduced (all-reduce) to finally update the model parameters synchronously. The memory footprint due to DP at each GPU is dependent on the minibatch size ($=$ batch size/DP-degree), sequence length, model dimension, model weights, gradients, and optimizer states. Tensor model parallelism alleviates the memory requirement of model related parameters. Essentially, TP partitions a tensor operation across multiple devices (e.g., the model weight matrix is split across rows or columns). Depending on the partitioning, each device might end up having the partial sums due to tensor contraction that need to be further reduced before going to the next stage of the computation, consequently causing network overhead. Here, the same data is copied to all participating devices. Using Megatron-LM's TP parallelization strategy~\cite{Shoeybi2019MegatronLMTM}, which is explained in Section~\ref{subsec:megatron}, the communication is of the all-reduce type, and the overhead is minimized. PP is a type of model parallelism that involves distributing the layers across multiple devices. Each device processes a set of layers and passes the activations to the next device. In PP, the minibatch is further divided into multiple microbatches that are passed in a pipeline fashion. This approach can introduce idle times, known as pipeline bubbles. This bubble time can be reduced using techniques like interleaved pipeline schedules as discussed in~\cite{Shoeybi2019MegatronLMTM}. 
TP distributes the attention and MLP blocks, but not the Dropout and Layer-norm following them. Despite being computationally inexpensive, the Dropout and Layer-norm contribute to a considerable amount of activation memory. Sequence parallelism was proposed to parallelize these blocks along the sequence dimension to proportionally reduce their memory footprint without incurring communication overhead. Since training is data intensive, usually combinations of DP, TP, SP, and PP are used to scale the performance across several hundreds or thousands of nodes. While for inference the data is limited and thus, most implementations involve only TP across a few devices within a node. Of course, batched inference with LLMs may require more devices to fit the model and data into the device memory. 

We consider all the above-discussed issues starting from the characteristics of the workloads in terms of task graphs, parallelization, mapping of that onto a system architecture, and modeling the performance of every kernel at each device and system level to derive key insights on performance. Our main contributions in this paper are given below: 
\begin{itemize}
\item We construct a general framework to model and analyze the performance of distributed LLM training and inference workloads that thoroughly considers the impacts of compute, memory sub-system, and network communication.
\item Upon extensive validation (GEMM, GEMV, training, and inference) and performance analysis, we bring in the insights behind the massive performance gain achieved by NVIDIA from A100 to B200.
\item We run design space exploration at different technology nodes and for different DRAM technologies to investigate the importance of compute vs memory.
\item We analyze the impact of off-chip memory technology scaling and network on LLM training and inference. Also, explore the GEMM level analysis of memory and compute boundedness for inference.  
\end{itemize}

\section{Related Work, Trends and Gap areas}
\label{sec:rel}

Numerous prior works~\cite{Timeloop, SMAUG, GAMMA, CoSA, Interstellar, MindMappings} focus on design space exploration for conventional Deep Neural Networks (DNNs). However, there are relatively few studies in the LLM space ~\cite{AMPeD, ASTRA-sim2.0, LLMViewer, DeepFlow, LLMCompass}.
In~\cite{Timeloop}, the authors propose an analytical model to assess the impact of different memory hierarchies on performance and energy consumption for convolution kernels. However, the model does not support end-to-end DNN performance modeling and focuses only on optimizing a single layer.
Ref.~\cite{SMAUG}, presents a DNN framework that can be simulated in a cycle-level SoC simulator, which incorporates data transformation, movement cost, and software framework overheads. 
In~\cite{GAMMA}, the authors propose a domain-specific genetic algorithm-based method for efficient DNN mappings considering a comprehensive map space including computation order, tile sizes, and parallelization.
The mapping proposed in~\cite{CoSA}, further improved the DNN computation scheduling problem by mapping the multiple scheduling problems into a single constrained optimization problem that can be solved directly without incurring the high cost of iterative scheduling.
The model proposed in~\cite{MindMappings}, utilizes a gradient-decent based algorithm to search the accelerator-cost function space in order to determine the optimum mapping. 

LLMs, nevertheless, have their own unique challenges that are different from DNNs. To optimize hardware utilization for LLM execution, different mapping strategies for compute-bound training and memory-bound inference are essential.
AMPeD~\cite{AMPeD} is an analytical model for end-to-end performance modeling of distributed transformer training that explores different parallelization strategies, tunes accelerator and system specifications. However, it neglects the modeling of the memory subsystem, technology implications, and network behavior.
In~\cite{ASTRA-sim2.0}, the authors emphasize on the communication network and scale-out modeling for server grade architecture. However, the authors do not study different mappings, architectural organizations, and technologies.
The analytical framework, in~\cite{LLMViewer}, offers a comprehensive review of efficient LLM inference and introduces a roofline-based analytical tool for systematic analysis. This framework identifies hardware bottlenecks and provides insights into memory and compute requirements for each layer of LLMs. However, it focuses on existing hardware and does not include customization of components like bandwidth, memory, and compute. Additionally, visibility into the GEMM operations within a layer is still a work in progress.
DeepFlow as proposed in~\cite{DeepFlow}, integrates micro architecture generation, compute graph transformation, device mapping, and performance prediction engine. It adopts a gradient descent technique to explore the design space with the knowledge of hardware specifications. The model characterizes only the LSTM workloads and validates with the existing GPUs.
In the paper~\cite{LLMCompass}, the authors introduces a hardware evaluation framework for LLM inference workloads. It is fast, accurate, and versatile, allowing for detailed evaluation of various hardware designs. The model breaks down GPU/TPU hardware to the systolic array level, offering greater design flexibility. It includes an automatic mapper for performance-optimal scheduling and an area-based cost model for architectural decision-making. However, it primarily focuses on inference and does not address LLM training.

To tackle the challenges of end-to-end performance modeling and workload analysis of LLM training and inference workloads, we propose a general analytical methodology or platform, \emph{Optimus}\footnote{Optimus is an in-house proprietary platform to perform early design space exploration of LLM workloads}. We conduct performance scaling analysis, explore the impact of logic and memory scaling on performance, and perform design space exploration at different semiconductor technology nodes to identify optimized system architectures. 
\section{Methodology}
\label{sec:model}
We build on DeepFlow's~\cite{DeepFlow} strong foundation and extend the framework extensively to model state-of-the-art LLM workloads, and architectures like advanced GPUs, TPUs, and future custom designs, enabling us to investigate the interplay between software and hardware to expose performance bottlenecks.
DeepFlow is a cross-stack pathfinding framework based on analytical performance modeling that integrates technology parameters and system-level architecture with workload characteristics like compute graphs and parallelization strategies. 
At the core, it applies a hierarchical roofline model with a memory-sub system aware tiling to predict the performance of GEMMs.
However, DeepFlow is currently tailored to older machine learning workloads, such as LSTMs, which can be reduced to a single bulky GEMM operation. Additionally, the framework's evaluation and validation have been conducted solely on old-generation GPU architectures, specifically P4 and V100. Significant efforts are needed to construct the configuration file for a new architecture as DeepFlow requires tedious low-level technology parameter specifications to derive important quantities like area per cell, energy per flip, etc.
This approach makes it difficult to model new-generation commercial architectures, like NVIDIA GPUs, since the technology details are generally not revealed.

\begin{figure}[b]
    \centering
    \includegraphics[width=0.95\linewidth]{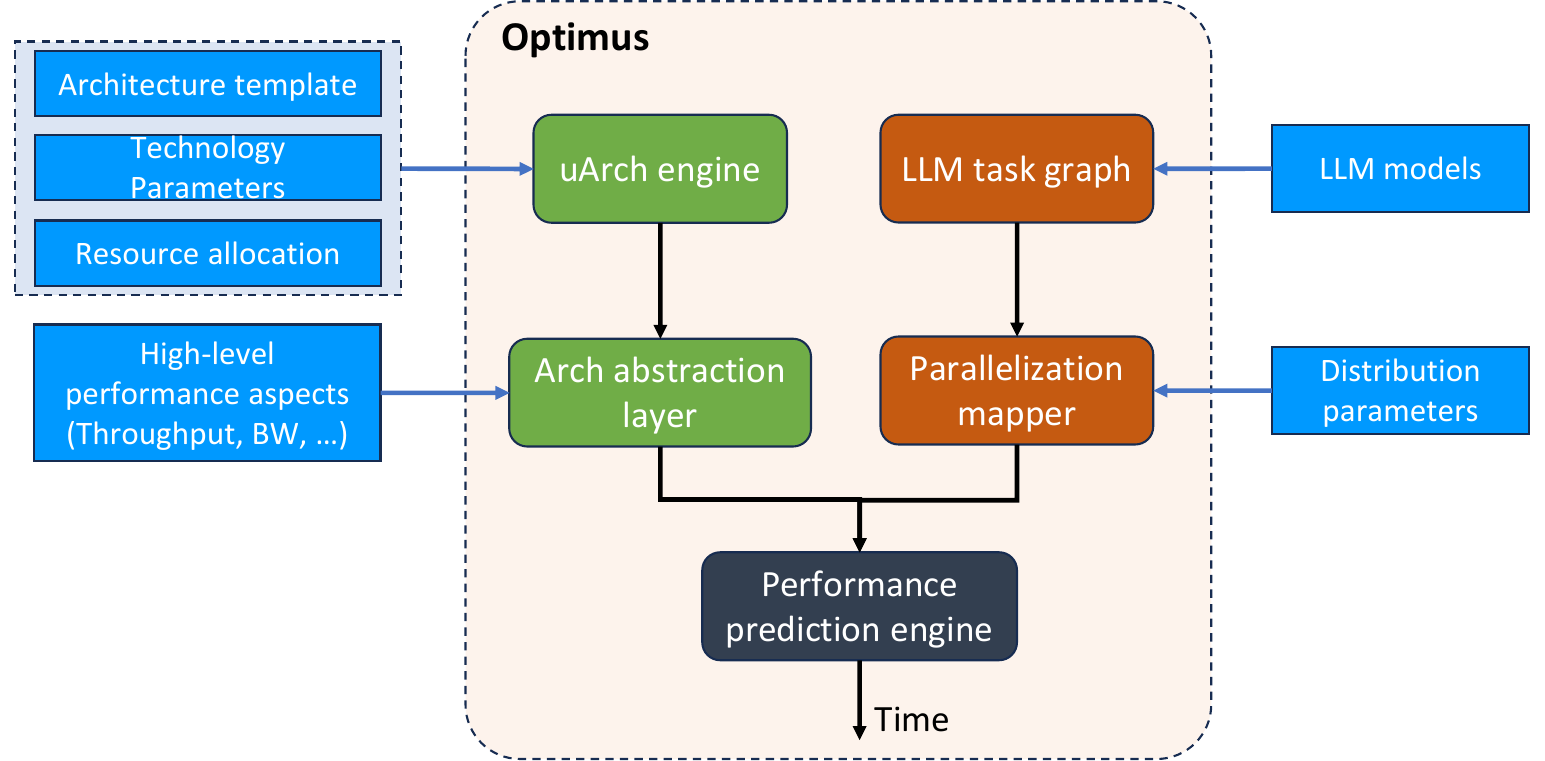}
    \caption{Overview of our performance modeling framework: $\mu$Arch engine generates a microarchitecture from the inputs. The architecture abstraction layer constructs a high-level representation of the underlying architecture. Given an LLM workload, the framework builds a task graph and parallelizes across multiple devices based on mapping. The performance prediction engine predicts the execution time.}
    \label{fig:optimus}
\end{figure}

\subsection{Framework overview}
Fig. \ref{fig:optimus} illustrates the high-level overview of our methodology. We start with the task graph of LLM training or inference and map that onto the system architecture based on the chosen parallelization strategy and mapping. We adopt the optimized mapping strategy proposed in the Megatron scheme~\cite{Shoeybi2019MegatronLMTM}. At every device level for GEMMs or GEMVs, we follow a hierarchical Roofline model based on DeepFlow~\cite{DeepFlow}. Different activation recomputation techniques are also incorporated to minimize model memory footprint, offering more optimization options as discussed later. 

On the system architecture side, we introduce an intermediate architecture abstraction layer between the microarchitecture engine and the performance prediction engine. This abstraction layer extracts the high-level performance drivers, such as compute throughput and memory sub-system bandwidths, derived from the underlying microarchitecture engine.
It can also directly receive a high-level system description from external inputs, which avoids tedious microarchicture parameter calibration and greatly eases the process of constructing the system configuration description for any new architectures without compromising prediction accuracy. This enables us to extend the studied processors to include modern GPUs (e.g., A100, H100, B100, and B200), TPUs, and custom architectures.


\subsection{Mapping \& Parallelization strategy}\label{subsec:megatron}
\begin{figure}[t]
    \centering
    \includegraphics[width=0.9\linewidth]{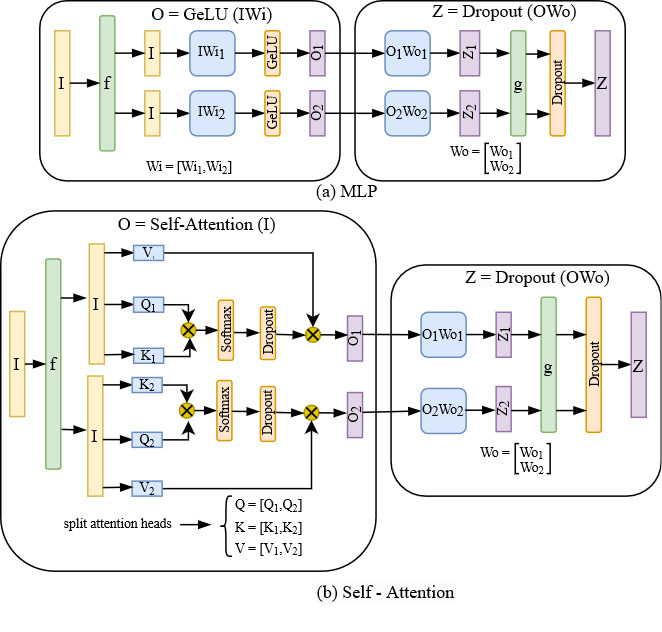}
    \caption{The model parallelism strategy proposed in Megatron-LM paper~\cite{Shoeybi2019MegatronLMTM} effectively reduces the need for synchronization and communication.}
    \label{fig:megatron}
\end{figure}

As discussed in Section~\ref{subsec:parallelization}, two primary parallelization strategies for scaling DNN training are data parallelism and model parallelism, with model parallelism further classified into tensor, pipeline and sequence parallelism~\cite{Shoeybi2019MegatronLMTM, Korthikanti2022ReducingAR}. We adopt the model parallelization strategy outlined in the Megatron paper~\cite{Shoeybi2019MegatronLMTM} for training LLMs. This approach focuses on minimizing synchronization requirements between processing devices by partitioning matrices effectively, thus reducing the need for communication. We also model the sequence parallelism proposed in \cite{Korthikanti2022ReducingAR} for distributing Dropout and Layer-norm layers.
For simplicity, we focus on explaining the concept in relation to the MLP block of a transformer layer, but as depicted in Figure~\ref{fig:megatron}, it applies to the MHA block as well.

Similar to any other MLP, a GEMM operation between the input matrix (\emph{I}) and the weight matrix (\emph{Wi}) is followed by a non-linear function, such as GELU. To parallelize the GEMM while minimizing synchronization needs, \emph{Wi} can be partitioned along its columns. This way, when the input is multiplied by each partition, the non-linear function can be applied independently without requiring synchronization. However, the resulting output (\emph{O}), which serves as the input to the second MLP layer, is also partitioned along its columns. Consequently, in the second MLP layer, the corresponding weight matrix (\emph{Wo}) should be divided along its rows in the same proportion as the previous layer's output. This ensures that corresponding elements from both matrices are on the same device and can be multiplied to produce partial results. The partial results of the second GEMM are then reduced across the processing devices before being passed to the dropout layer. This approach splits both GEMMs in the MLP block across processing devices and requires only a single all-reduce operation in the forward pass per MLP block. It is worth mentioning that in the MHA block, computation of \emph{Key}, \emph{Query}, and \emph{Value} for different attention heads are processed in parallel across multiple devices since they are independent. The rest follows a similar parallelization strategy as the MLP block. Besides, we also adopt various PP schedules, including GPipe \cite{gpipe}, PipeDream-Flush \cite{pmlr-v139-narayanan21a}, and Interleaved 1F1B \cite{eff_megatron}.
\subsection{Activation recomputation}
Training LLMs demands a large amount of memory. The total required memory mainly consists of model parameters, optimizer states, and activations. Since training dictates that the activations in all layers need to be saved for backward gradient computation, activations become the critical bottleneck for scaling LLMs. We have implemented two mainstream activation recomputation techniques to alleviate the storage issue, namely 
full recomputation \cite{Chen2016TrainingDN} and selective recomputation \cite{Korthikanti2022ReducingAR}. 
Full recomputation checkpoints LLM layers and recompute all the activations by executing the forward pass again. Despite saving substantial memory space, it doubles the forward pass time. The required activation memory size, $A_{\rm full}$, can be expressed as
\begin{equation}
    A_{\rm full} = N_{\rm ckp}A_{\rm inp} + L/N_{\rm ckp}(A_{\rm tot} - A_{\rm inp})
\end{equation}
where $N_{\rm ckp}$ is the number of checkpoints, $A_{\rm inp}$ is the input activation size of a transformer layer, $A_{\rm tot}$ is the total activation size of a transformer layer, and $L$ is the number of layers.

Selective recomputation selects the memory-intensive parts of LLM layers, such as softmax and dropout outputs, which are not computationally expensive for recomputation. The required activation memory size, $A_{\rm sel}$, can be expressed as,
\begin{equation}
    A_{\rm sel} = L(A_{\rm tot} - (A_{\rm sm} + A_{\rm do\_\rm mask} + A_{\rm do\_\rm out}))  
\end{equation}
where $A_{\rm sm}$ is the size of the input activation to a Softmax layer, $A_{\rm do\_\rm mask}$ is the mask size of the dropout layer after the Softmax layer, $A_{\rm do\_\rm out}$ is the size of the output from the Dropout layer.

\subsection{Modeling All-to-All communication}
Both training and inference use all-gather or all-reduce communication collectives where all devices within a node or the full system communicate with each other for gathering or reducing data. The communication involves fetching the data from the respective device memory and sending it in a pipeline fashion to hide the latency. 
We model ring-topology~\cite{ring_reduce} and Double binary trees topology for global communications~\cite{double_tree, nvidia_tree}. 
Ring all-reduce is a bandwidth-optimal algorithm, suitable for data-intensive communication when the impact of latency is negligible \cite{PATARASUK2009117}. The communication cost is determined by the slowest connection between processors, independent of the number of processors. The algorithm consists of two stages: scatter-reduce and all gather. All the processors are arranged in a logical ring. In the scatter-reduce stage, each processor sends a chunk of data to its right neighbor and reduces the data received from its left neighbor. Each processor ends up with one chunk of the final result. An all-gather operation is then performed across all the processors. The data transfer communication time, $T_r$, can be expressed as,
\begin{equation}
    T_r = \frac{2K}{N\times \rm BW}(N-1) + 2~l~(N-1)
\end{equation}
where $K$ is the data volume to be transferred, $N$ is the number of processors, $\rm BW$ is the network bandwidth between processors, $l$ is the network latency. While during training the latency term is negligible, for inference, its contribution can be non-negligible due to low data volume. Thus, we also model double binary trees-based communication that is both bandwidth and latency optimal~\cite{double_tree}. The revised communication time is expressed as~\cite{Rabenseifner}, 
\begin{equation}
    T_r = \frac{2K}{N\times \rm BW}(N-1) + 2~l~\log_2(N)
\end{equation}
The second term in the above reduces the impact of latency and helps scale inference up to 8 GPUs. It is worth noting that for inference, the data volume is generally low and the network bandwidth is underutilized. We apply a utilization factor to derive the actual bandwidth.

\subsection{KV-cache modeling}\label{subsec:KV_cache}
Unlike training, GenAI inference processes one token at a time (in the auto-regressive generation or decoding phase) sequentially depending on previously generated tokens~\cite{efficient_inf}. This sequential nature complicates the use of parallelization techniques, such as flash-attention~\cite{flashattention2}, that are effective during training. 
As mentioned in Section~\ref{subsec:Transformer}, \emph{Key}s and \emph{Value}s are used to calculate the scaled dot-product attention for each \emph{Query}. During the decoding phase of an LLM, the attention calculation for previous tokens is repeated at each generation step.

KV-caching optimizes this process by focusing solely on calculating attention for the new token while caching the previously computed \emph{Key}s and \emph{Value}s. 
The required total KV-cache size is given by $(2 \times {\rm batch~size} \times {\rm context} \times {\rm precision} \times {\rm \#layers} \times {\rm embedding~dimesnion})$. The first factor of 2 is due to the \emph{Key} and \emph{Value}-matrices. Without KV-cache the \emph{Key}s and \emph{Value}s for all the previous tokens need to be recomputed. 
The trade-off for this approach is the increased memory and bandwidth required to store and load the \emph{Key} and \emph{Value} states.







\subsection{Design space exploration framework}
Design Space Exploration (DSE) relies on connecting technology parameters to an architecture template following a system topology to create a micro-architecture that serves as a blueprint for the underlying device. With a given budget and allocation of hardware resources (i.e., area, power, and chip perimeter) on the micro-architecture, the DSE framework derives the essential coarse-grained quantities, such as compute throughput, memory capacities, memory bandwidths, and network bandwidths (both inter-node and intra-node). The analytical performance model estimates the performance (i.e., execution time) of a given workload based on these high-level quantities that define the system. Thus, the DSE framework solves a constrained optimization problem. The search space contains all possible choices of area, power, and perimeter fractions for each component in the micro-architecture. The constraint is a given resource budget. A gradient-descent search algorithm is employed to find the optimal design point that minimizes the execution time.



\section{Validation}
\label{sec:val}
Extensive validations of the proposed methodology have been conducted through experimental measurements and comparison with published data across various platforms. Below we present detailed results for GPUs.

\subsection{Distributed GEMM and GEMV validation}\label{subsec:GEMM_val}
The primitive component of LLM workloads is the GEMM kernel. Thus, it is essential to validate its implementation in different scenarios. 
The training phase generally involves fat GEMMs. Whereas, skinny GEMMs such as GEMV, dominate computation in the inference phase due to the autoregressive token generation phase. This section covers the validation of both types.

\begin{figure}
    \centering
    \includegraphics[width=0.8\linewidth]{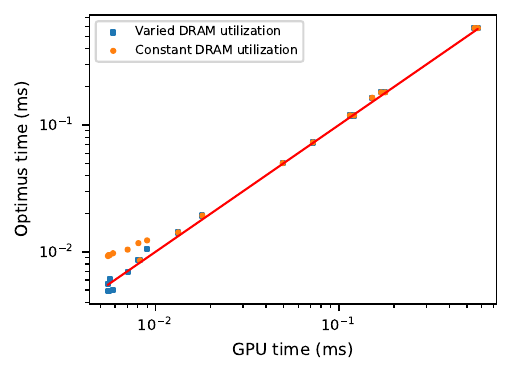}
    \caption{Correlation between GPU runtime and our prediction for GEMV validation on a single A100 GPU.}
    \label{fig:gemv_A100}
\end{figure}

Fat GEMM kernels are generally compute-bound kernels due to high arithmetic intensity-- they were intensively validated on the old-generation GPUs in DeepFlow. We extend the validation studies and verified its prediction accuracy on advanced GPUs (A100s). On the other hand, GEMV kernels are typically bounded by the memory bandwidth in GPUs. Since GEMV kernels move small volumes of data between memory levels, DRAM bandwidth is usually underutilized. The utilization depends largely on the matrix/vector dimension. In order to capture the effect of the underutilized memory bandwidth, a memory bandwidth utilization factor is introduced 
to adjust the roofline model-based time calculation.
Matrix/vector dimensions were selected to cover a wide range of kernel types used in the LLMs. The selected GEMV kernels are profiled on A100 GPUs, their memory utilization in multilevel memory as well as compute utilization are recorded and clustering techniques are used to quantify the DRAM utilization factors. 
This process results in reducing the absolute percentage error between the measurements and the prediction to 5.4\% (depicted in blue in Fig. \ref{fig:gemv_A100}). 
We also simplify the modeling process by applying a constant DRAM utilization factor to all the GEMV kernels, resulting in negligible errors for large matrices (indicated by orange points). For smaller sizes, the software overhead has a non-negligible impact.  Worth mentioning that we have extended our modeling framework to accommodate TPUs and custom architectures.

\subsection{LLM training validation for GPUs}
Our framework provides training time predictions for decoder-based transformer models, such as GPTs. Running large-scale LLM training is not feasible for us, thus, we have validated our framework and methodology against the published data~\cite{Shoeybi2019MegatronLMTM, Korthikanti2022ReducingAR}. We model the impact of various parallelism strategies, including TP, DP, PP, and SP. Due to high memory consumption by activations, different activation recomputation techniques are considered, including full recomputation and selective recomputation. Table~\ref{tab:megatron_validation} shows the comparisons between the reported training time per batch for different systems of A100 GPUs (8 to 3072 GPUs) and our predictions for different parallelism settings across various GPT models (22 B - 1 T parameters models). Among all the cases, the relative errors are mostly well below 10\%. It is worth noting that TP and SP are always implemented within a node due to their higher communication overhead. DP and PP are usually implemented across nodes. For PP, the interleaved scheduling strategy assigns multiple pipeline stages in each device and allows for a large number of microbatches with a small memory footprint, which minimizes the pipeline bubbles. Our framework accurately models every aspect of different parallelisms and can also produce accurate memory breakdown.


\begin{table*}[t]
\caption{Validation of training time per batch for LLMs on systems of A100 GPUs with varying choices of parallelism strategies for different GPT models.}
\label{tab:megatron_validation}
\begin{tabular}{|cccccccc|}
\hline
\multicolumn{1}{|c|}{Model}     & \multicolumn{1}{c|}{\# GPUs} & \multicolumn{1}{c|}{Batch size} & \multicolumn{1}{c|}{DP-TP-PP-SP} & \multicolumn{1}{c|}{Activation recomputation} & \multicolumn{1}{c|}{$t_{ref}$ in s from \cite{Shoeybi2019MegatronLMTM}\cite{Korthikanti2022ReducingAR}} & \multicolumn{1}{c|}{$t_{pred}$ in s} & $\delta E$ ~(\%) \\ \hline
\multicolumn{8}{|l|}{Only TP and PP}                                                                                                                                                                                                                                              \\ \hline
\multicolumn{1}{|c|}{GPT-22B}   & \multicolumn{1}{c|}{8}       & \multicolumn{1}{c|}{4}          & \multicolumn{1}{c|}{1-8-8-1}     & \multicolumn{1}{c|}{full}                     & \multicolumn{1}{c|}{1.4}               & \multicolumn{1}{c|}{1.4}              & 2.1        \\ \hline
\multicolumn{1}{|c|}{GPT-175B}  & \multicolumn{1}{c|}{64}      & \multicolumn{1}{c|}{64}         & \multicolumn{1}{c|}{1-8-8-1}     & \multicolumn{1}{c|}{full}                     & \multicolumn{1}{c|}{18.1}              & \multicolumn{1}{c|}{16.9}             & 6.9        \\ \hline
\multicolumn{1}{|c|}{GPT-530B}  & \multicolumn{1}{c|}{280}     & \multicolumn{1}{c|}{280}        & \multicolumn{1}{c|}{1-8-35-1}    & \multicolumn{1}{c|}{full}                     & \multicolumn{1}{c|}{49.1}              & \multicolumn{1}{c|}{46.8}             & 4.6        \\ \hline
\multicolumn{1}{|c|}{GPT-1008B} & \multicolumn{1}{c|}{512}     & \multicolumn{1}{c|}{512}        & \multicolumn{1}{c|}{1-8-64-1}    & \multicolumn{1}{c|}{full}                     & \multicolumn{1}{c|}{94.4}              & \multicolumn{1}{c|}{87.9}             & 6.9        \\ \hline
\multicolumn{8}{|l|}{TP, PP and SP}                                                                                                                                                                                                                                               \\ \hline
\multicolumn{1}{|c|}{GPT-22B}   & \multicolumn{1}{c|}{8}       & \multicolumn{1}{c|}{4}          & \multicolumn{1}{c|}{1-8-8-8}     & \multicolumn{1}{c|}{selective}                & \multicolumn{1}{c|}{1.1}               & \multicolumn{1}{c|}{1.1}              & 0.0        \\ \hline
\multicolumn{1}{|c|}{GPT-175B}  & \multicolumn{1}{c|}{64}      & \multicolumn{1}{c|}{64}         & \multicolumn{1}{c|}{1-8-8-8}     & \multicolumn{1}{c|}{selective}                & \multicolumn{1}{c|}{13.8}              & \multicolumn{1}{c|}{12.9}             & 5.9        \\ \hline
\multicolumn{1}{|c|}{GPT-530B}  & \multicolumn{1}{c|}{280}     & \multicolumn{1}{c|}{280}        & \multicolumn{1}{c|}{1-8-35-8}    & \multicolumn{1}{c|}{selective}                & \multicolumn{1}{c|}{37.8}              & \multicolumn{1}{c|}{35.5}             & 6.2        \\ \hline
\multicolumn{1}{|c|}{GPT-1008B} & \multicolumn{1}{c|}{512}     & \multicolumn{1}{c|}{512}        & \multicolumn{1}{c|}{1-8-64-8}    & \multicolumn{1}{c|}{selective}                & \multicolumn{1}{c|}{71.5}              & \multicolumn{1}{c|}{69.1}             & 3.4        \\ \hline
\multicolumn{8}{|l|}{DP, TP and PP}                                                                                                                                                                                                                                               \\ \hline
\multicolumn{1}{|c|}{GPT-310B}  & \multicolumn{1}{c|}{1920}    & \multicolumn{1}{c|}{2160}       & \multicolumn{1}{c|}{15-8-16-1}   & \multicolumn{1}{c|}{full}                     & \multicolumn{1}{c|}{37.6}              & \multicolumn{1}{c|}{34.1}             & 9.5        \\ \hline
\multicolumn{1}{|c|}{GPT-530B}  & \multicolumn{1}{c|}{2520}    & \multicolumn{1}{c|}{2520}       & \multicolumn{1}{c|}{9-8-35-1}    & \multicolumn{1}{c|}{full}                     & \multicolumn{1}{c|}{54.2}              & \multicolumn{1}{c|}{51.2}             & 5.5        \\ \hline
\multicolumn{1}{|c|}{GPT-1008B} & \multicolumn{1}{c|}{3072}    & \multicolumn{1}{c|}{3072}       & \multicolumn{1}{c|}{6-8-64-1}    & \multicolumn{1}{c|}{full}                     & \multicolumn{1}{c|}{102.4}             & \multicolumn{1}{c|}{100.7}            & 1.6        \\ \hline
\end{tabular}
\centering
\end{table*}

\begin{table*}[h]
\caption{Validation (with~\cite{nvidia_inference}) of inference latency on systems of A100 and H100 GPUs with varying degree of TP for three different Llama models. Here, the batch size is set to 1, summarization and generation stages involve 200 tokens.}
\label{tab:inference_validation}
\begin{tabular}{|c|c|c|c|c|c|c|c|c|}
\hline
Model & \# GPUs & TP & t$_{\rm nvidia}$ in ms (A100) & t$_{\rm pred}$~in ms (A100)& $\delta E$~(\%) & t$_{\rm nvidia}$ in ms (H100)  & t$_{\rm pred}$ in ms (H100) & $\delta E$ ~(\%)\\ \hline
Llama2-70B  & 8   & 8  & 4735 & 4284 & 9.5 & 3202 & 3147 & 1.7 \\ \hline
Llama2-70B  & 4   & 4  & 6403 & 6019 & 6.0 & 4116 & 3986 & 3.2 \\ \hline
Llama2-70B  & 2   & 2  & 10500 & 10042 & 4.4 & 6267 & 6186 & 1.3 \\ \hline
Llama2-13B  & 8   & 8  & 1693 & 1514 & 10.6 & 1201 & 1209 & 0.7 \\ \hline
Llama2-13B  & 4   & 4  & 1894 & 1748 & 7.7 & 1431 & 1258 & 12.1 \\ \hline
Llama2-13B  & 2   & 2  & 2499 & 2492 & 0.2 & 1717 & 1617 & 5.8 \\ \hline
Llama2-13B  & 1   & 1  & 3884 & 4263 & 9.7 & 2396 & 2599 & 8.5 \\ \hline
Llama2-7B  & 8   & 8  & 1187 & 1096 & 7.7 & 828 & 899 & 8.6 \\ \hline
Llama2-7B  & 4   & 4  & 1280 & 1166 & 8.9 & 924 & 869 & 5.6 \\ \hline
Llama2-7B  & 2   & 2  & 1544 & 1526 & 1.2 & 1143 & 1016 & 11.1 \\ \hline
Llama2-7B  & 1   & 1  & 2190 & 2472 & 12.9 & 1440 & 1522 & 5.7 \\ \hline
\end{tabular}
\centering
\end{table*}
\subsection{Inference validation for GPUs}
Since inference involves skinny GEMM or GEMV, we first validate the model by running them on A100 GPU as discussed in Section~\ref{subsec:GEMM_val}. We validate the inference latencies reported by NVIDIA in~\cite{nvidia_inference} with the predicted values using our performance model for Llama-2-7b, Llama2-13b, and Llama2-70b on A100-80GB and H100-SXM GPU systems (Table~\ref{tab:gpt3}). For each LLM model and a given GPU type, we validate the strong scaling from 1 to 8 GPUs. Here, the batch size is set to 1, and the prefill and generation involve 200 tokens. In all the cases, we match the actual reported numbers within a relative error of 13\%. The scaling in performance from A100 to H100 is largely due to the change in DRAM technology from HBM2e (bandwidth of 1.9 TBPs) to HBM3 (bandwidth of 3.35 TBPs). Interesting to note that, inference scales poorly with the number of GPUs, unlike training due to the sequential nature of token generation in the prediction phase that suffers from relatively less compute requirement and is purely memory-bound. The only anomaly we see is in the case of the Llama2-7B parameter model on 8 H100 GPUs: while scaling from 4 to 8 GPUs, the predicted inference latency goes up while NVIDIA reports otherwise. Although the gain reported by NVIDIA is very modest, we see an opposite trend. This is primarily due to the absence of a rigorous network simulator in our performance model which could give a more realistic utilization of the network communication bandwidth based on the data volume. 

\section{Case studies: Training}
Different aspects of LLM training workload analysis, such as memory profiling, performance projection, and bottleneck analysis, can be conducted through the 
Optimus framework. In this section, we performed case studies to profile the LLM training memory footprint, project GPU performance scaling across multiple GPU generations (i.e., from A100 to the latest B200), and assess the performance scaling of semiconductor technology node from 12 nm (N12) to 1 nm (N1). Here, the technology node refers to a specific generation of manufacturing process technology.

\begin{figure}[b]
    \centering
    \includegraphics[width=\linewidth]{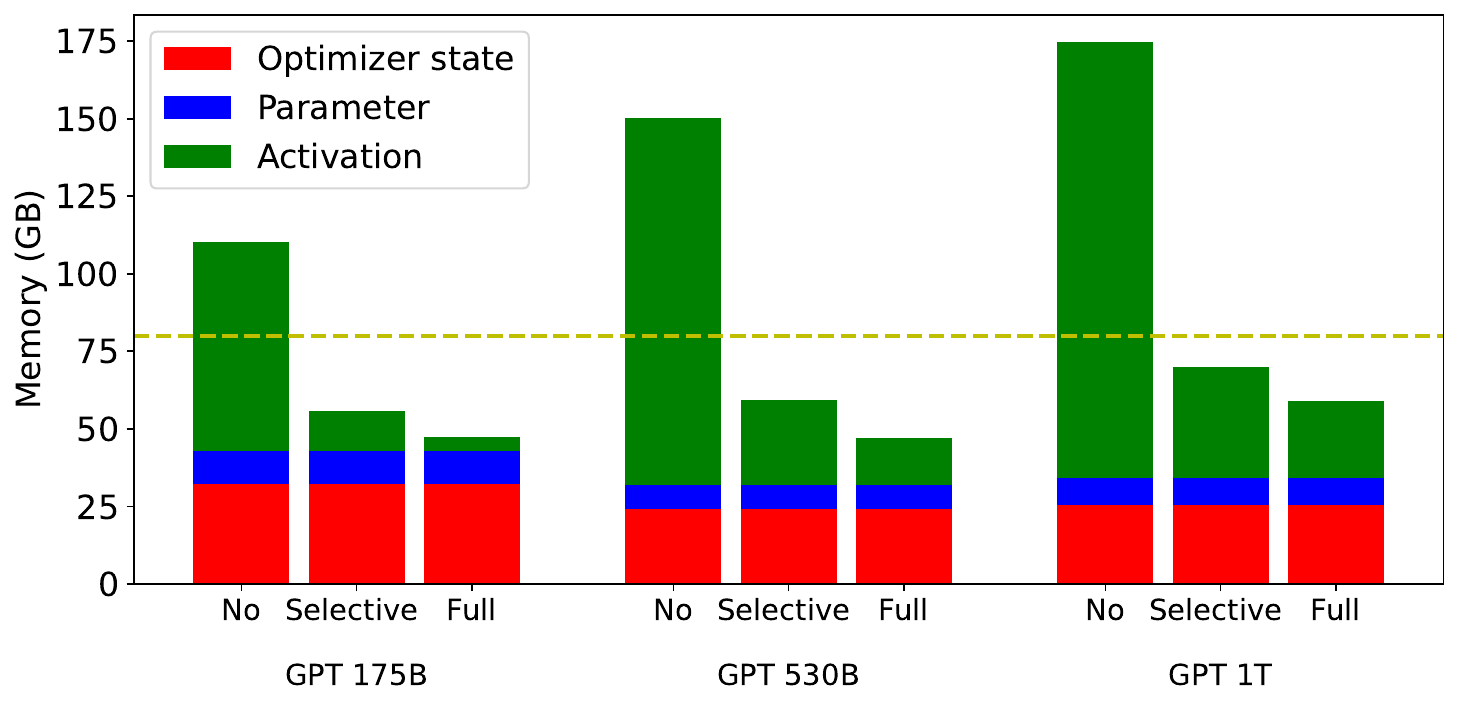}
    \caption{Memory breakdown for training GPT models. The dash line indicates the NVIDIA A100 memory capacity, 80 GB. For each GPT model, three activation recomputation methods are compared: no recomputation, selective recomputation, and full recomputation.}
    \label{fig:mem_breakdown}
\end{figure}

\subsection{Memory dissection}
One of the challenges in training billion to trillion-parameter LLMs is memory overflow. Data parallelism and model parallelism reduce memory proportionally but suffer from fundamental scaling limitation. It is important to know if an LLM can fit in the device memory of a system before performing any other analysis. The choice of parallelism mapping strategy largely depends on the resulting model memory footprint. Memory profiling helps identify the bottleneck, so that we can apply the corresponding memory optimization techniques. We can also determine the best parallelism mapping or training settings for an LLM model on a certain hardware system or the suitable hardware system for desired settings.

As an example, we performed memory profiling on three different GPT models in three activation recomputation scenarios: no recomputation, selective recomputation, and full recomputation. The training configurations are available in TABLE \ref{tab:megatron_validation}. We consider mixed-precision training with 2 bytes. Fig. \ref{fig:mem_breakdown} shows the memory breakdown. We can clearly observe the difference in memory footprint made by the choice of the recomputation technique. With no recomputation, an LLM can not generally fit in the device memory unless a very small batch size or large degrees of parallelism are applied, which largely degrades the training efficiency. Selective recomputation exhibits a small difference from the full recomputation and causes very little computational overhead. The extra memory space can be utilized for further optimizing training efficiency.

\subsection{Projected GPU performance scaling}
\begin{figure}
    \centering
    \includegraphics[width=0.85\linewidth]{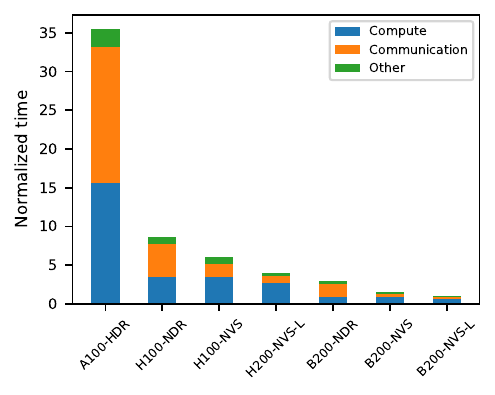}
    \caption{Training performance scaling across multiple GPU generations for GPT-3 175B. Training times are normalized against that of B200-NVS-L. The A100 cluster is connected through HDR InfiniBand network, while the others are configured with NDR infiniBand network or NVLink switch system (NVS). L indicates a larger batch size. {\em Other} includes weight update time + pipeline bubble time.}
    \label{fig:gpu-scaling}
\end{figure}

With the dire demand for accelerated processing of AI workloads, NVIDIA GPUs have experienced many generations of improvement, providing massive compute and scalability. Using our performance model, we project the training time for the GPT3-175B model run on various GPU systems, namely A100, H100, H200, and B200. Table \ref{tab:gpt3} lists the configurations used for the projection. The inter-node communication in the A100 cluster is through HDR InfiniBand (IB) network (200 GB/s), while the nodes in the more advanced GPU clusters are connected through the NDR IB network (400 GB/s) or NVLink switch system (NVS). Fig. \ref{fig:gpu-scaling} reveals a clear trend of performance scaling across these generations. H200-NVS-L and B200-NVS-L used a larger batch size of 4096 to demonstrate their large DRAM capacity. Improved over A100, H100 triples the compute throughput and introduces an FP8 transformer engine for mixed-precision training. H100-NDR gives rise to around 4x speedup. An additional factor of 2 can be obtained by using the NVLink Switch system. Due to a larger DRAM capacity, H200 can accommodate an even larger batch size and hence, further accelerate the training by 3x. The latest breakthrough from NVIDIA, B200, enables FP4 processing and further boosts the performance by 3x with NDR IB and by 14x with NVS. B200 also features a larger DRAM capacity and faster memory bandwidth. Using a larger batch size leads to 12x more acceleration. Our projections from A100 to H100 and from H100 to B200 are aligned with the reported data from NVIDIA \cite{andersch2022nvidia, nvidia2024dgxb200}. 

The scaling of compute capacity and communication speed from one generation to another is not at the same rate. Through performance modeling, we found that the LLM training configuration, such as the degrees of all the parallelisms, can largely affect the scaling trend since it affects the ratio between compute time and communication time. However, in general, training configurations are not disclosed by NVIDIA. Through our performance model, we can obtain all the details and gain insights into how the significant performance gain is achieved and how to optimally utilize the computing power and network throughput of a GPU.


\subsection{Technology node scaling}
\begin{figure}[t]
    \centering
    \includegraphics[width=0.85\linewidth]{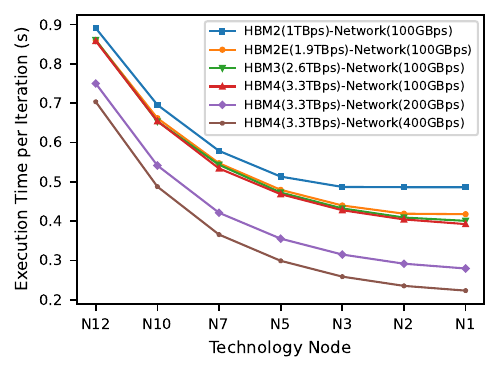}
    \caption{Scaling technology node for different memory and network types.}
    \label{fig:tech-scaling}
\end{figure}

The Optimus framework creates a full-stack platform linking semiconductor technology specifics with the performance predictions of LLMs. This facilitates the analysis of how advancements in low-level technology nodes impact model performance, offering insights into performance bottlenecks and future trends for better HW-SW codesign. This case study showcases training performance evolution for an LLM with the advancement in technology development and projects the performance of future nodes. Here, we consider the GPT-7B parameter model distributed across 1024 GPUs. The training configuration is displayed in TABLE \ref{tab:case_studies}. Seven logic technology nodes were explored, ranging from N12 to N1. With technology logic scaling, transistors effectively get smaller, allowing for more of them to fit onto a chip, thus, enhancing the compute density. We followed the assumption of iso-performance scaling between consecutive nodes for an optimistic prediction \cite{STILLMAKER201774, DeepFlow}, which determined the scaling factors of 1.8 and 1.3 for area and power, respectively. Four generations of HBM technology (HBM2 (1TB/s), HBM2E (1.9TB/s), HBM3(2.6TB/s), and HBM4 (projected 3.3TB/s)) and three types of inter-node IB network technology (NDR-x8 (100 GB/s), XDR-x8 (200GB/s) and GDR-x8 (400GB/s)) were considered. We apply the DSE method to optimize the architecture at each technology node based on the area and power budget. The scaling results are shown in Fig. \ref{fig:tech-scaling}.
\begin{table}[t]
\caption{Training configurations of case studies for different GPT models. Here, total \#GPUs = DP $\times$ TP $\times$ PP.}
\label{tab:case_studies}
\begin{tabular}{|c|c|c|c|c|}
\hline
Model & Batch size & Seq length & Vocab size & DP-TP-SP-PP \\ \hline
GPT-175B & 1024/4096  & 2048            & 51200      & 128-8-8-8   \\ \hline
GPT-7B & 512  & 2048            & 51200      & 64-4-4-4   \\ \hline
\end{tabular}
\label{tab:gpt3}
\vspace{-10pt}
\end{table}
\begin{figure*}[ht!]
\centering
    \begin{subfigure}{0.31\linewidth}
        \centering
        \includegraphics[width=\linewidth]{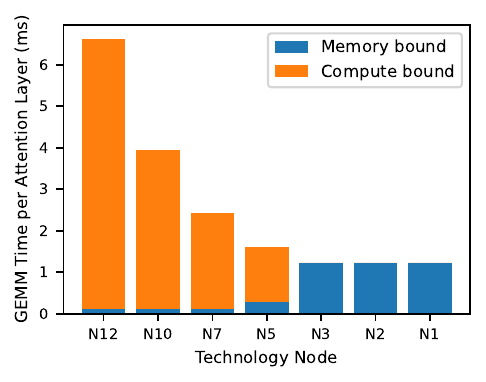}
        \caption{HBM2-NDR}
        \label{fig:HBM2}
    \end{subfigure}
    \begin{subfigure}{0.31\linewidth}
        \centering
        \includegraphics[width=\linewidth]{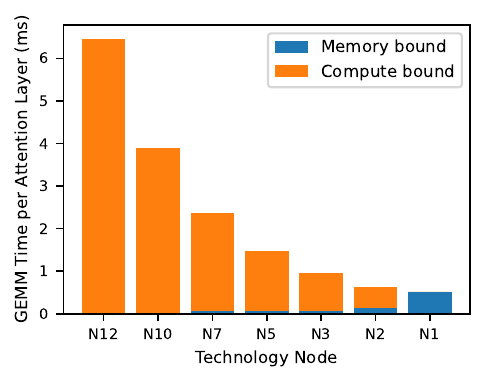}
        \caption{HBM3-NDR}
        \label{fig:HBM3}
    \end{subfigure}
    \begin{subfigure}{0.31\linewidth}
        \centering
        \includegraphics[width=\linewidth]{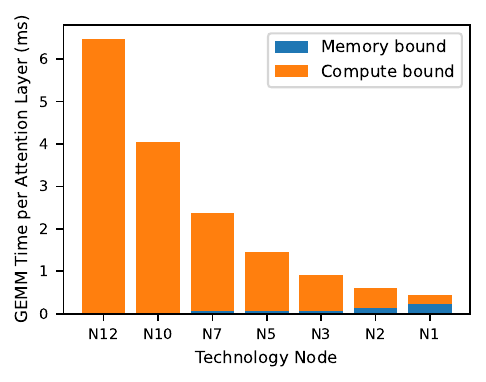}
        \caption{HBM4-NDR}
        \label{fig:HBM4}
    \end{subfigure}    
    \caption{GEMM time breakdown for a single transformer layer in terms of bound types for different HBM technologies: (a) HBM2, (b) HBM3 and (c) HBM4. The GEMM times were extracted from the corresponding technology node scaling experiments.}
    \label{fig:gemm_breakdwon_scaling}
\end{figure*}
The node scaling improves the core throughput, cache capacity, and bandwidth. The training time decreases drastically initially and then tends to saturate at very advanced nodes beyond N5. During training phase, transformer layers are generally compute bound. However, when the compute throughput of the device improves significantly with the scaling, these layers start becoming bounded by device memory, like DRAM. At this point, the node scaling hardly contributes to the performance improvement. Fig. \ref{fig:gemm_breakdwon_scaling} shows the breakdown of bound types in one attention layer. The impact of memory boundedness becomes dominant gradually with the scaling. Using more advanced HBM technology, such as moving from HBM2 to HBM2E, leads to a significant runtime reduction. Further improvement with HBM3/4 is not observed because the model performance becomes bounded by the network bandwidth. It is worth noting that DeepFlow framework predicted that the model performance was bounded by L2 cache instead of compute or device memory, which is not aligned with actual experiments \cite{Ivanov2020DataMI}\cite{flashattention}. 

Logic node scaling and memory technology advancement reduce device kernel time. However, in distributed training, the communication overhead between compute nodes grows substantially at the same time due to parallelism. The performance scaling is particularly constrained by the slow inter-node network. Enhancing network bandwidth from 100 GB/s to 400 GB/s markedly improves training times. Pushing for faster networks has been the focus of industries. For example, NVIDIA's NVLink Switch System has transformed inter-node networking to match intra-node network performance, resulting in significant performance gains.
\begin{table}[t]
\caption{Identification of performance bottlenecks at different matrix multiply functions per transformer layer in the summarization phase of inference. The data are Llama2-13B model, single A100 and H100 systems with half precision, B = 1, and summarization of 200 tokens.}
\scalebox{0.87}{
\centering
  \begin{tabular}{|l|cc|cc|}
  \hline
    \multirow{2}{*}{GEMM-function} &
      \multicolumn{2}{c|}{A100} &
      \multicolumn{2}{c|}{H100} \\
      & {t ($\mu$s)} & {bound-type} & {t ($\mu$s)} & {bound-type} \\
      \hline
    merged-head X.W$_{K/Q/V}$= K,Q,V & 82 & {\rm compute} & 32 & {\rm memory} \\
    single head Q.K$^{\rm T}$ = R & 3 & {\rm memory} & 2 & {\rm memory}\\
    single head softmax(R).V = Z & 3 & {\rm memory} & 2 & {\rm memory}\\
    Z.W=O & 42 & {\rm compute} & 17 & {\rm memory}\\
    O.W$_{\rm MLP1}$ = O$_1$ & 216 & {\rm compute} & 81 & {\rm memory}\\
    O$_1$.W$_{\rm MLP2}$ = O$_2$ & 109 & {\rm compute} & 42 & {\rm memory}\\
    \hline
  \end{tabular}}
   \label{table:comp-mem_infernce}
\end{table}
\begin{figure}
\vspace{-10pt}
\centering
    \includegraphics[width=0.75\linewidth]{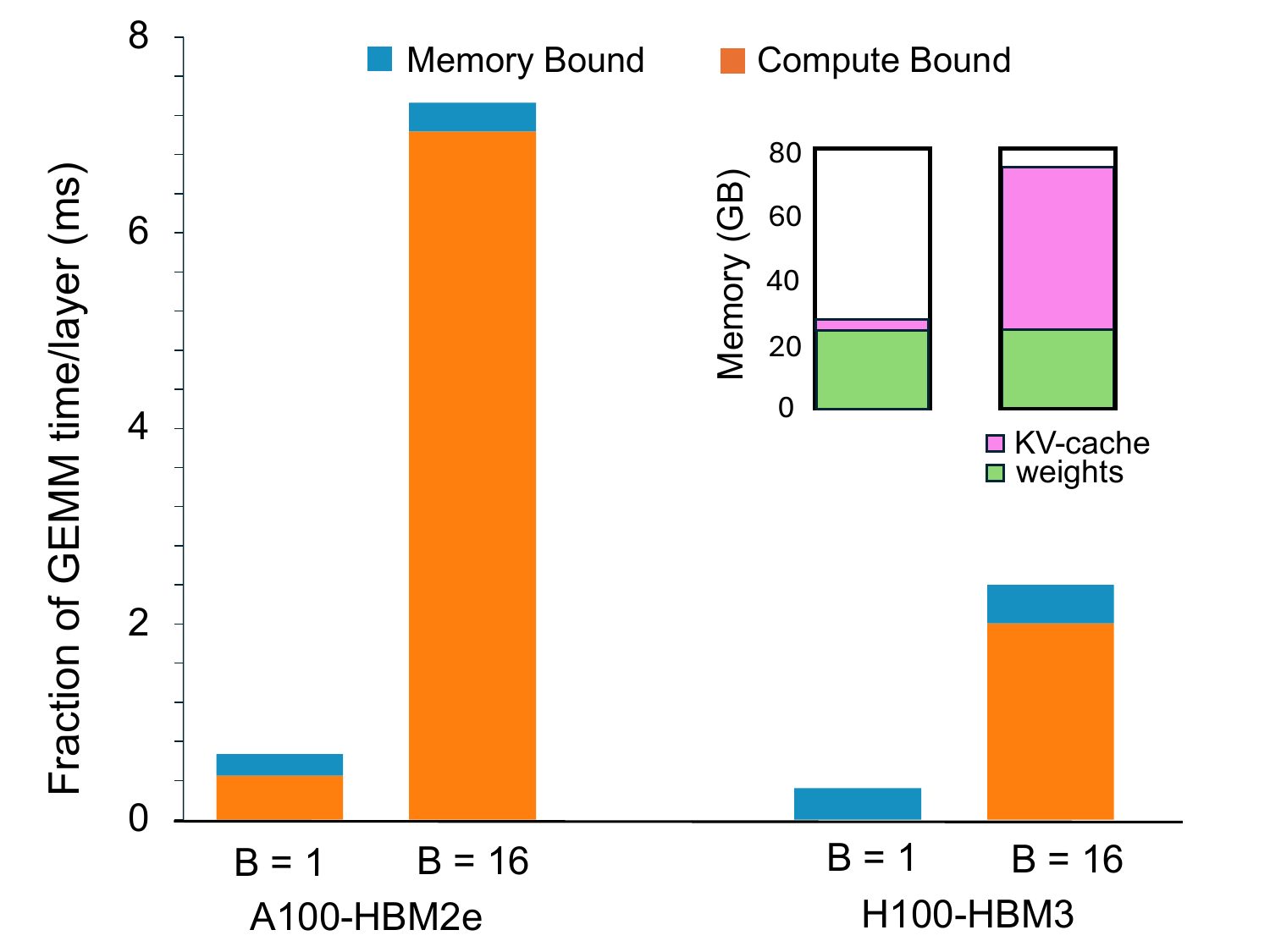}
    \caption{GEMM time breakdown per layer in terms of bound types during the summarization phase of inference for the batch size of 1 and 16. The generation phase is completely memory bound. The inset shows the memory capacities, required memory size for kv-cache and model weights.}
    \label{fig:comp-mem_infernce}
\end{figure}
\section{Case studies: Inference}
In this section, we primarily investigate the importance of compute versus memory throughput on LLM inference. 
\subsection{Memory boundedness of inference}
The autoregressive generation phase in inference is typically DRAM bandwidth bound for both A100 and H100 GPUs even when batch size (B) $> 1$ (for half-precision). However, the prefill or summarization phase can be compute bound depending on the accelerator type, precision, batch size, and the prompt length. In Table~\ref{table:comp-mem_infernce}, we analyze the performance bottleneck of all the GEMM kernels in the summarization phase of inference using the Llama2-13B model with half precision. We identify that on A100, the following GEMMs, single head QK$^T$ = R and softmax(R)V = Z are particularly DRAM memory bandwidth bound due to the limited reuse of the bytes transferred from the device memory. The other GEMMs are compute bound due to their shape and dimensions. On H100, all the GEMMs in both prefill and generation phases are DRAM-bound (see Table~\ref{table:comp-mem_infernce}). This implies that as the compute scales, performance for inference becomes completely determined by the memory technology. However, it is worth noting that at low precision, the memory transaction volume decreases and the compute throughput goes up, thus impacting the arithmetic intensity of the GEMM-- these options are being exploited in NVIDIA GPUs. In Fig.~\ref{fig:comp-mem_infernce}, we show the GEMM time breakdown for a single layer based on the compute versus memory boundedness for two different batch sizes, B = 1 and 16. On A100, when B=1, roughly 67\% of the time is spent on compute-bound GEMMs -- this percentage grows to 96\% when B = 16. Whereas, for H100, the compute dominated time fraction reduces to zero when B = 1, but grows to 85\% when B = 16. Larger batch sizes, thus, improve inference throughput but at the cost of latency. However, the growth of latency with B is rather modest. The inset shows the memory footprint of the KV-cache and the model weights for the Llama2-13B model when B = 1 and 16.  
\begin{figure*}[h!]
    \centering
\includegraphics[width=0.87\linewidth]{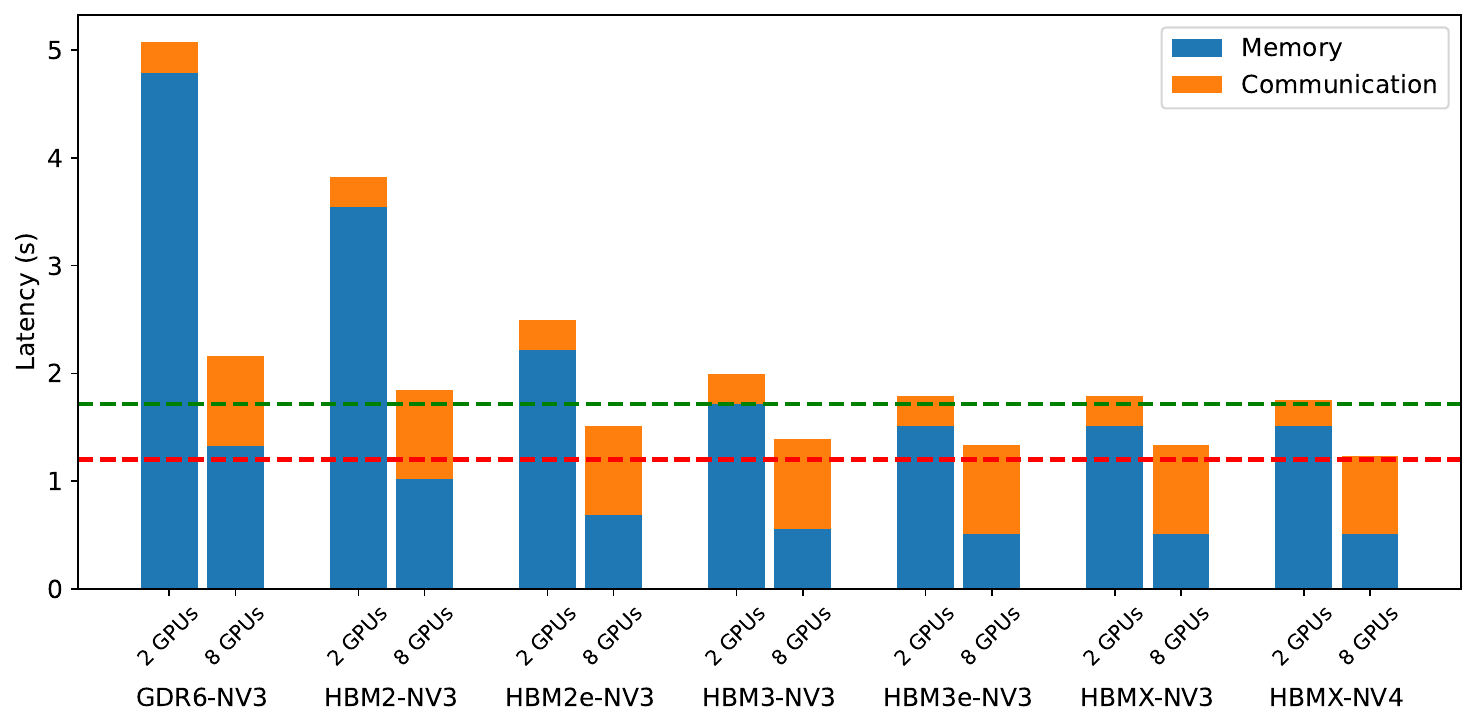}
    \caption{Impact of memory technology scaling on inference latency as predicted by the performance model. The data are for the Llama2-13B parameter model, batch size = 1, prefill and generation of 200 tokens. The on-chip specifications are same as A100. The horizontal dashed lines correspond to 2xH100-HBM3 and 8xH100-80GB-HBM3 connected via NVLink-Gen4.}
    \label{fig:memory-scaling}
\end{figure*}
\subsection{Memory technology scaling}
Since the performance of inference is primarily bound by off-chip memory, we investigate the impact of DRAM memory technology scaling on the inference latency for multi-GPU systems using our performance model. For this case study, we again consider the Llama2-13B parameter model (kv-cache and weights fit into GPU's device memory), batch size = 1, prefill and generation of 200 tokens. We keep the compute fixed at the 7 nm technology node (A100) and vary the DRAM technology from GDR6 (bandwidth = 600GB/s) to HBM3e (bandwidth = 4.8TB/s). Further, we consider a futuristic memory technology, HBMX with the peak bandwidth of 6.8 TB/s. For all these cases, the network is considered as NVLink-Gen3 (NV3). Lastly, we also consider the case of A100-HBMX GPUs connected by NVLink-Gen4 (NV4) interconnect. The results for 2-GPU and 8-GPU systems are presented in Fig.~\ref{fig:memory-scaling}. We see that up to HBM3, the performance almost scales linearly with the DRAM bandwidth and it slows down from HBM3 to HBM3e. Beyond that, no such performance gain in memory time is observed since the DRAM bandwidth surpasses the last level cache bandwidth and the problem starts to become L2-bound. Further speed-up can be obtained by scaling the on-chip memory bandwidth and capacity or the intra-node network. While changing the network from NVLink-Gen3 (NV3) to NVLink-Gen4 (NV4) a modest gain in communication $\sim 12\%$ is observed. The horizontal dashed lines correspond to the latency of 2xH100-HBM3e and 8xH100-HBM3e systems as references. Note that at HBM3e, H100 system is slightly faster than the projected A100-HBM3e system-- that's primarily due to the faster on-chip memory (no difference in DRAM technology) and faster network (NV4). It is also worth noting that although the compute throughput of H100 (5 nm) for half-precision is 989.4 TFLOPS, more than 3x of A100's compute throughput, it does not help improve inference performance further, implying that in principle, one can use an older technology node with the same memory technology without degrading the performance at the same precision. Another important aspect while running inference on a 4 or 8-GPU system, is the network communication overhead. We identify that for 8 GPUs, communication time is roughly 1.6x of memory time (for Llama2-13B). Solutions like optical networks can really alleviate this communication bottleneck\cite{aakash2024}.


\section{Conclusion}

In this work, we analyze LLM training and inference workloads within an analytical performance modeling framework. We describe the modeling approach, extensively validate it with published data, and analyze different test cases. In particular, we investigate the impact of compute and memory technology scaling on the performance of LLM training and inference. At a single device level, the performance model is based on a hierarchical roofline model, and the communication across multiple devices (or nodes) is derived using the Megatron-LM-like mapping. This combination leads to remarkably accurate performance predictions at scale both for training and inference. Further improvements can be made with a first principle approach of determining realistic memory bandwidth utilization for memory-bound problems like, inference and by complementing it with a network simulator to incorporate the impact of network traffic and congestion. Our future investigation includes evaluating future compute system architectures and technologies for LLM training and inference, integrating a cost and an energy model into the current performance modeling framework, and performing complete performance per TCO analysis.

\bibliographystyle{IEEEtranS}
\bibliography{reference}

\end{document}